\documentclass{ifacconf}

\usepackage{graphicx}      
\usepackage{natbib}        

\usepackage{dsfont}
\newcommand{\ones}{\mathds{1}}

\usepackage[toc,page]{appendix}

\usepackage{amsmath}
\usepackage{amsfonts}
\usepackage{algorithm}
\usepackage{xcolor}
\usepackage{balance} 

\usepackage{paralist}
\usepackage{enumitem}

\usepackage{algorithm}
\usepackage[noend]{algpseudocode}
\usepackage{subcaption}
\usepackage{epstopdf}
\usepackage{balance}
\usepackage{epstopdf}
\usepackage{float}

\usepackage{tikz} 
\usepackage{pgfplots}

\theoremstyle{plain} 
\newtheorem{theorem}{Theorem}
\newtheorem{lemma}{Lemma}
\newtheorem{corollary}{Corollary}

\theoremstyle{definition} 

\newtheorem{problem}{Problem}
\newtheorem{assumption}{Assumption}

\theoremstyle{remark} 
\newtheorem{remark}{Remark}

\definecolor{mycolor1}{rgb}{0.00000,0.44700,0.74100}%
\definecolor{mycolor2}{rgb}{0.85000,0.32500,0.09800}%
\definecolor{mycolor3}{rgb}{0.92900,0.69400,0.12500}%
\definecolor{mycolor4}{rgb}{0.49400,0.18400,0.55600}%
\usepackage{filecontents}

\begin{document}
\begin{frontmatter}

\title{Cooperative System Identification \\ via Correctional Learning\thanksref{footnoteinfo}} 

\thanks[footnoteinfo]{This work supported by the Wallenberg AI, Autonomous Systems and Software Program
        (WASP), the Swedish Research Council and the Swedish Research Council Research Environment NewLEADS under contract 2016-06079.}

\author[First]{In\^{e}s Louren\c{c}o} 
\author[First]{Robert Mattila} 
\author[First]{Cristian R. Rojas}
\author[First]{Bo Wahlberg}

\address[First]{Division of Decision and
        Control Systems, School of Electrical Engineering and Computer Science, KTH Royal Institute of Technology, Stockholm, Sweden. \\(e-mails: \{ineslo, rmattila, crro, bo\}@kth.se).}

\begin{abstract}                

We consider a cooperative system identification scenario in which an expert agent (teacher) knows a correct, or at least a good, model of the system and aims to assist a learner-agent (student), but cannot directly transfer its knowledge to the student. For example, the teacher's knowledge of the system might be abstract or the teacher and student might be employing different model classes, which renders the teacher's parameters uninformative to the student. 
In this paper, we propose \emph{correctional learning} as an approach to the above problem: Suppose that in order to assist the student, the teacher can intercept the
observations collected from the system and modify them to maximize the amount of information
the student receives about the system.
We formulate a general solution as an optimization problem, which for a multinomial system instantiates itself as an integer program. Furthermore, we obtain finite-sample results on the improvement that the assistance from the teacher results in (as measured by the reduction in the variance of the estimator) for a binomial system.
In numerical experiments, 
we illustrate the proposed algorithms and verify the theoretical results that have been derived in the paper. 
\end{abstract}


\end{frontmatter}

\section{INTRODUCTION}
\label{sec:intro}

System identification concerns the calibration and validation of models of dynamical systems from observed data \citep{ljung1998system}. An important component is experiment design and, in particular, optimal input/excitation design in system identification for control \citep{5717863, 7879927, gerencser2009identification, wittenmark1975stochastic}. Pre-processing of the measured output signals is of utmost importance, for example for outlier detection \citep{hodge2004survey}. Similarly, controlled sensing deals with applications where it is possible to use multiple sensors with different qualities and also costs. Which sensor should the decision-maker choose at each time instant to provide the next measurement? This problem is also referred to as the sensor scheduling problem, the measurement control problem, sensor-adaptive signal processing or the active sensing problem. We refer to \cite[Ch.~8]{krishnamurthy_2016} for a thorough overview of this field.

The process of system identification from experiment design to the resulting model estimate can be very costly and time consuming. It often relies on a priori information about the underlying system, for example in a Bayesian setting, \citep{PETERKA198141}. The objective of this paper is to study an alternative approach how external knowledge can be incorporated in processing the data for identification. More specifically, we assume that an expert (\emph{teacher}) has knowledge about the underlying system and aims to assist a \emph{student} (i.e., the system identification procedure). The central question in this paper is:
\begin{center}
\emph{    How can the teacher assist (e.g., accelerate or make more accurate) the data-driven learning process of the student?}
\end{center}
A challenge is that it may be impossible for the expert to directly transfer its knowledge to the student. For example, the expert's knowledge might be abstract (consider for example teaching someone how to drive a car), or the expert and student might operate in different model classes and/or parametrizations, which renders the expert's parameters meaningless to the student. Moreover, the model might simply be too complex to be transmitted. For example, the GTP-3 model \citep{brown2020language} integrates on the order of 100 billion parameters, which translates to a memory requirement of around 350 GB for a trained GPT-3 model. Transmission of such a huge parameter set is non-trivial due to constraints on real-time performance specifications and communication constraints.
Direct communication between the expert and student might also be restricted due to privacy concerns, or in a defense setting, transmission of a trained model between two powers can be diplomatically prohibited.

In this paper, we propose a novel approach called \textit{correctional learning}. In it, the expert assists the student by intercepting and modifying the data collected from the system, in a way such that the student is better able to estimate a model of the system than with the raw unmodified data sequence.
For example, in the defense setting alluded to above, correctional learning provides means of stealthily transmitting a model by intercepting and modifying, e.g., radar measurements. Correctional devices for error codes have a long history in information theory \cite[Section~XII]{shannon1948mathematical} but, to the best of our knowledge, these ideas are novel for system identification. 

To summarize, the main contributions of this paper are:
\begin{itemize}
\item We propose \emph{correctional learning} as a means to perform cooperative system identification;
\item We derive both general and specific algorithms for correctional learning. These are formulated as optimization problems that minimize the discrepancy between the student's empirical probability distribution over the observation set, and that induced by the model known to the teacher. The problems take into account a limited budget of the teacher;
\item We analyze the reduction in variance of the student's estimator as a function of data and budget size. In Corollary 1, we provide guarantee that the teacher is successfully able to transfer its knowledge using correctional learning;
\item The proposed algorithms and theoretical results are illustrated and evaluated in numerical simulations with promising results.
\end{itemize}

The rest of the paper is structured as follows. Section~\ref{sec:preliminaries} first introduces notation and preliminaries, and then provides a general formulation of cooperative system identification. Section \ref{sec:algs} gives algorithms for general correctional learning for cooperative system identification. Section \ref{sec:examples} derives specific algorithms and results for bi- and multinomial distributions. Finally, Section \ref{sec:results} validates the framework in numerical experiments. Full proofs and extended simulations are available in \citep{lourenco_arXiv_2020}.

\section{Preliminaries and Problem Formulation}
\label{sec:preliminaries}

In this section, we define our notation and introduce the cooperative system identification problem formally: the student collects data and performs standard system identification, and the teacher, who has access to a good model, aims to accelerate the learning process.


\subsection{Notation}

All vectors are column vectors (unless transposed) and inequalities between
vectors are considered element-wise. The element $i$ of a vector $v$ is $[v]_i$. A generic probability density (or mass) function is denoted as $p(\cdot)$. The vector of ones is denoted as $\ones$, the set of non-negative real numbers as $\mathbb{R}^+_0$, and the indicator function as $I(\cdot)$. 
The $\ell_1$ and $\ell_2$ norms of a vector $v$ are denoted $\|v\|_1$ and $\|v\|_2$, respectively.

\subsection{Student: System identification}
\label{sec:formulation}


The student performs standard system identification. That is, it samples observations from the system and uses these to estimate a model. 
More formally, assume for simplicity that time $k$ is discrete. We consider that the data $\mathcal{D}$ is given by observations $y_k \in \mathcal{Y}$, where $\mathcal{Y}$ is the observation space, collected during $N$ time-steps: $\mathcal{D} = \{y_k\}_{k=1}^N$. 

We assume that the system producing the observations is given by some model $m_0 \in \mathcal{M}$, where $\mathcal{M}$ is a model class. Moreover, we assume that the model class is parametric, so that $m \in \mathcal{M}$ can be equivalently characterized by a vector $\theta \in \mathbb{R}^p$ of some $p$ parameters. Let $\theta_0$ denote the parameters corresponding to the true model $m_0$.

Then, each observation is sampled according to
\begin{equation}
    y_k \sim p(y | y_{k-1}, \dots, y_1; \theta_0),
    \label{eq:observations}
\end{equation}
for $k = 1, \dots, N$.

In order to estimate a model, the student uses a measure of fit $F(m, \mathcal{D})$ that measures how well a model $m$ describes observed data $\mathcal{D}$:
\begin{equation}
    \hat{m} \in \arg \min_{m \in \mathcal{M}} F(m, \mathcal{D}),
    \label{eq:min_goodness_of_fit_model}
\end{equation}
where $\hat{m}$ is the estimated model.

A typical choice for the goodness-of-fit $F$ is the likelihood-function of the observed data.
Assuming a parametric model class, we can equivalently formulate \eqref{eq:min_goodness_of_fit_model} as
\begin{equation}
    \hat{\theta} \in \arg \min_{\theta \in \Theta} F(m(\theta), \mathcal{D}),
    \label{eq:theta}
\end{equation}
where $\Theta \subset \mathbb{R}^p$ is the feasible parameter set.

Note that, in general, the student aims to estimate a model from some model class $\mathcal{M}'$ that is different from the true model class: $\mathcal{M}' \neq \mathcal{M}$. As we will discuss below, this prevents the teacher from directly transmitting knowledge of $m_0$ (since $m_0$, or its associated parameters, are meaningless to the student: $m_0 \not\in \mathcal{M}'$).


Under suitable assumptions on assumptions (on $\mathcal{M}$ and $F$), the estimator $\hat{m}$ will converge to $m_0$ (or the closest point in $\mathcal{M}'$) -- see, e.g., \cite{ljung1978convergence} for details.


\subsection{Teacher: Cooperative system identification}

In the cooperative scenario that we investigate, the teacher knows the true model $m_0$, or equivalently $\theta_0$, and aims to convey this information to the student.

As mentioned above, if $\mathcal{M}' \neq \mathcal{M}$ (that is, if the teacher and student employ different model classes) then giving $m_0$ (or $\theta_0$) to the student is meaningless. This also holds true if the model class is shared $\mathcal{M}' = \mathcal{M}$, but different parametrizations are used by the student and teacher (since then, $\theta_0$ is incomprehensible to the student). For clarity and to simplify the theoretical analysis later on, we make the following assumption:\\

 \begin{assumption}[Shared Model Class]
    The teacher's and student's model classes, $\mathcal{M}$ and $\mathcal{M}'$, respectively, coincide: $\mathcal{M}' = \mathcal{M}$.
    \label{ass:model_class}
\end{assumption}
Nevertheless, the algorithms generalize with straight-forward modifications and we believe the results we obtain can form a basis for a more detailed analysis in future work.

The cooperative system identification problem is:\\




\begin{problem}(Cooperative System Identification).
Observations $y_k$ are being sampled from system $m(\theta_0) \in \mathcal{M}$, as in \eqref{eq:observations}. A teacher knows $\theta_0$, but is, for various reasons, unable to communicate $\theta_0$ directly to the student, who is forming an estimate according to \eqref{eq:theta}. Provide a method for the teacher to improve the learning process of the student.
\label{pr:cl}
\end{problem}

The last sentence in Problem \ref{pr:cl} (in particular, the word ``improve'') merits some discussion. Under suitable assumptions, the estimator computed in problem \eqref{eq:theta} is consistent: as the number of data points $N$ grows to infinity, $\theta_0$ will be reconstructed with probability one. Here, then, with ``improve'' we mean reducing the asymptotic covariance of the student's estimator. Similarily, for a finite $N$, improving the learning process of the student should be interpreted as minimizing the variance of the estimator formed using $N$ samples.\\ 

\begin{remark}
Problem \ref{pr:cl} can be reinterpreted in an adversarial setting where the expert aims to hinder the learning process of the student. 
This is common for example in privacy scenarios, where the expert has knowledge about the system but does not want another agent to be able to estimate it exactly \citep{Showkatbakhsh2016system,bottegal2017preserving,lourencco2020protect}. 
The algorithms we discuss in the remaining part of the paper apply equally to this setting by, essentially, flipping a sign.
\end{remark}


\section{Cooperative System Identification via Correctional Learning}
\label{sec:algs}




The cooperative system identification problem (Problem \ref{pr:cl}) can be approached in several ways. For example, in Bayesian system identification \citep{PETERKA198141}, the teacher can help formulating the student's prior -- this, however, requires that the student and teacher share parametrization. 
Similarly, the student could employ inverse filtering \cite{mattila2020} to obtain parameters without an explicit transfer of such information taking place.

Instead, in this section, we present the novel \emph{correctional learning} framework as a means to solve Problem \ref{pr:cl}: We allow the teacher to modify (``correct'') the observations seen by the student. That is, to adjust some observations in the original dataset $\mathcal{D} = \{y_k\}_{k=1}^N$ so that the student sees $\tilde{\mathcal{D}} = \{\tilde{y}_k\}_{k=1}^N$ instead. By doing this, information is communicated agnostically to the requirement of a shared model class and/or parametrization, as discussed below.

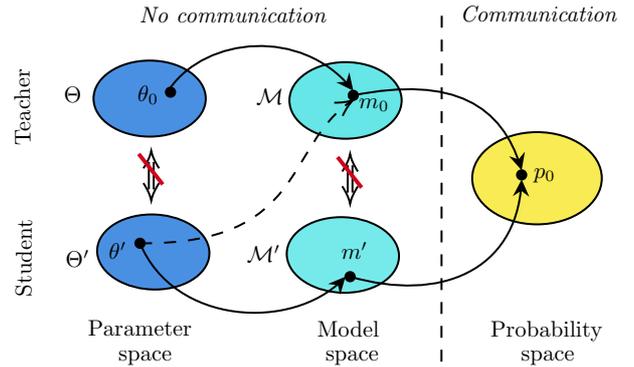
\begin{figure}
    \centering

\tikzset{every picture/.style={line width=0.75pt}} 

\begin{tikzpicture}[x=0.75pt,y=0.75pt,yscale=-0.8,xscale=0.8, every node/.style={scale=0.85}]

\draw  [fill={rgb, 255:red, 74; green, 144; blue, 226 }  ,fill opacity=1 ] (220,68.08) .. controls (220,54.87) and (235.67,44.17) .. (255,44.17) .. controls (274.33,44.17) and (290,54.87) .. (290,68.08) .. controls (290,81.29) and (274.33,92) .. (255,92) .. controls (235.67,92) and (220,81.29) .. (220,68.08) -- cycle ;
\draw  [fill={rgb, 255:red, 246; green, 230; blue, 32 }  ,fill opacity=0.77 ] (456.17,118.17) .. controls (456.17,102.15) and (474.26,89.17) .. (496.58,89.17) .. controls (518.9,89.17) and (537,102.15) .. (537,118.17) .. controls (537,134.18) and (518.9,147.17) .. (496.58,147.17) .. controls (474.26,147.17) and (456.17,134.18) .. (456.17,118.17) -- cycle ;
\draw  [fill={rgb, 255:red, 78; green, 226; blue, 226 }  ,fill opacity=0.86 ] (342.17,69.25) .. controls (342.17,55.95) and (357.84,45.17) .. (377.17,45.17) .. controls (396.5,45.17) and (412.17,55.95) .. (412.17,69.25) .. controls (412.17,82.55) and (396.5,93.33) .. (377.17,93.33) .. controls (357.84,93.33) and (342.17,82.55) .. (342.17,69.25) -- cycle ;
\draw    (382,66.06) .. controls (445.7,53.26) and (474.72,73.28) .. (486.31,110.84) ;
\draw [shift={(487,113.17)}, rotate = 254.12] [fill={rgb, 255:red, 0; green, 0; blue, 0 }  ][line width=0.08]  [draw opacity=0] (10.72,-5.15) -- (0,0) -- (10.72,5.15) -- (7.12,0) -- cycle    ;
\draw  [fill={rgb, 255:red, 74; green, 144; blue, 226 }  ,fill opacity=1 ] (292.22,163.97) .. controls (293.27,177.01) and (278.47,187.83) .. (259.16,188.15) .. controls (239.85,188.46) and (223.35,178.16) .. (222.31,165.12) .. controls (221.26,152.09) and (236.06,141.27) .. (255.37,140.95) .. controls (274.68,140.63) and (291.18,150.94) .. (292.22,163.97) -- cycle ;
\draw  [fill={rgb, 255:red, 73; green, 228; blue, 228 }  ,fill opacity=0.75 ] (410.47,165.84) .. controls (411.52,178.96) and (396.72,189.86) .. (377.42,190.18) .. controls (358.11,190.49) and (341.6,180.11) .. (340.55,166.99) .. controls (339.5,153.87) and (354.3,142.97) .. (373.6,142.65) .. controls (392.91,142.34) and (409.41,152.72) .. (410.47,165.84) -- cycle ;
\draw    (249,159.06) .. controls (270.56,210.94) and (343.98,214.06) .. (375.25,182.06) ;
\draw [shift={(377.11,180.06)}, rotate = 491.43] [fill={rgb, 255:red, 0; green, 0; blue, 0 }  ][line width=0.08]  [draw opacity=0] (10.72,-5.15) -- (0,0) -- (10.72,5.15) -- (7.12,0) -- cycle    ;
\draw    (257.5,108) -- (257.5,125.17)(254.5,108) -- (254.5,125.17) ;
\draw [shift={(256,132.17)}, rotate = 270] [color={rgb, 255:red, 0; green, 0; blue, 0 }  ][line width=0.75]    (10.93,-4.9) .. controls (6.95,-2.3) and (3.31,-0.67) .. (0,0) .. controls (3.31,0.67) and (6.95,2.3) .. (10.93,4.9)   ;
\draw [shift={(256,101)}, rotate = 90] [color={rgb, 255:red, 0; green, 0; blue, 0 }  ][line width=0.75]    (10.93,-4.9) .. controls (6.95,-2.3) and (3.31,-0.67) .. (0,0) .. controls (3.31,0.67) and (6.95,2.3) .. (10.93,4.9)   ;
\draw [color={rgb, 255:red, 208; green, 2; blue, 27 }  ,draw opacity=1 ][line width=1.5]    (248.19,102.89) -- (263.19,122.06) ;
\draw  [fill={rgb, 255:red, 0; green, 0; blue, 0 }  ,fill opacity=1 ] (265.11,64.06) .. controls (265.11,62.46) and (266.4,61.17) .. (268,61.17) .. controls (269.6,61.17) and (270.89,62.46) .. (270.89,64.06) .. controls (270.89,65.66) and (269.59,66.95) .. (268,66.95) .. controls (266.4,66.95) and (265.11,65.65) .. (265.11,64.06) -- cycle ;
\draw  [fill={rgb, 255:red, 0; green, 0; blue, 0 }  ,fill opacity=1 ] (379.11,66.06) .. controls (379.11,64.46) and (380.4,63.17) .. (382,63.17) .. controls (383.6,63.17) and (384.89,64.46) .. (384.89,66.06) .. controls (384.89,67.66) and (383.59,68.95) .. (382,68.95) .. controls (380.4,68.95) and (379.11,67.65) .. (379.11,66.06) -- cycle ;
\draw    (268,61.17) .. controls (290.66,29.48) and (349.21,23.19) .. (380.59,61.39) ;
\draw [shift={(382,63.17)}, rotate = 232.34] [fill={rgb, 255:red, 0; green, 0; blue, 0 }  ][line width=0.08]  [draw opacity=0] (10.72,-5.15) -- (0,0) -- (10.72,5.15) -- (7.12,0) -- cycle    ;
\draw  [fill={rgb, 255:red, 0; green, 0; blue, 0 }  ,fill opacity=1 ] (484.11,116.06) .. controls (484.11,114.46) and (485.4,113.17) .. (487,113.17) .. controls (488.6,113.17) and (489.89,114.46) .. (489.89,116.06) .. controls (489.89,117.66) and (488.59,118.95) .. (487,118.95) .. controls (485.4,118.95) and (484.11,117.65) .. (484.11,116.06) -- cycle ;
\draw  [fill={rgb, 255:red, 0; green, 0; blue, 0 }  ,fill opacity=1 ] (246.11,159.06) .. controls (246.11,157.46) and (247.4,156.17) .. (249,156.17) .. controls (250.6,156.17) and (251.89,157.46) .. (251.89,159.06) .. controls (251.89,160.66) and (250.59,161.95) .. (249,161.95) .. controls (247.4,161.95) and (246.11,160.65) .. (246.11,159.06) -- cycle ;
\draw  [fill={rgb, 255:red, 0; green, 0; blue, 0 }  ,fill opacity=1 ] (377.11,180.06) .. controls (377.11,178.46) and (378.4,177.17) .. (380,177.17) .. controls (381.6,177.17) and (382.89,178.46) .. (382.89,180.06) .. controls (382.89,181.66) and (381.59,182.95) .. (380,182.95) .. controls (378.4,182.95) and (377.11,181.65) .. (377.11,180.06) -- cycle ;
\draw    (486.82,122.27) .. controls (483.12,175.86) and (436.95,196.66) .. (380,180.06) ;
\draw [shift={(487,118.95)}, rotate = 92.04] [fill={rgb, 255:red, 0; green, 0; blue, 0 }  ][line width=0.08]  [draw opacity=0] (10.72,-5.15) -- (0,0) -- (10.72,5.15) -- (7.12,0) -- cycle    ;
\draw  [dash pattern={on 4.5pt off 4.5pt}]  (249,159.06) .. controls (352.95,160.16) and (339.29,97.44) .. (380.73,69.78) ;
\draw [shift={(382,68.95)}, rotate = 507.67] [color={rgb, 255:red, 0; green, 0; blue, 0 }  ][line width=0.75]    (10.93,-4.9) .. controls (6.95,-2.3) and (3.31,-0.67) .. (0,0) .. controls (3.31,0.67) and (6.95,2.3) .. (10.93,4.9)   ;
\draw    (381.5,110) -- (381.5,127.17)(378.5,110) -- (378.5,127.17) ;
\draw [shift={(380,134.17)}, rotate = 270] [color={rgb, 255:red, 0; green, 0; blue, 0 }  ][line width=0.75]    (10.93,-4.9) .. controls (6.95,-2.3) and (3.31,-0.67) .. (0,0) .. controls (3.31,0.67) and (6.95,2.3) .. (10.93,4.9)   ;
\draw [shift={(380,103)}, rotate = 90] [color={rgb, 255:red, 0; green, 0; blue, 0 }  ][line width=0.75]    (10.93,-4.9) .. controls (6.95,-2.3) and (3.31,-0.67) .. (0,0) .. controls (3.31,0.67) and (6.95,2.3) .. (10.93,4.9)   ;
\draw [color={rgb, 255:red, 208; green, 2; blue, 27 }  ,draw opacity=1 ][line width=1.5]    (372.19,104.89) -- (387.19,124.06) ;
\draw  [dash pattern={on 4.5pt off 4.5pt}]  (437.17,16) -- (437,237) ;

\draw (215,206) node [anchor=north west][inner sep=0.75pt]  [font=\normalsize] [align=left] {\begin{minipage}[lt]{50.341216pt}\setlength\topsep{0pt}
Parameter
\begin{center}
space
\end{center}

\end{minipage}};
\draw (358,206) node [anchor=north west][inner sep=0.75pt]  [font=\normalsize] [align=left] {\begin{minipage}[lt]{30.507928000000003pt}\setlength\topsep{0pt}
Model
\begin{center}
space
\end{center}

\end{minipage}};
\draw (466,206) node [anchor=north west][inner sep=0.75pt]  [font=\normalsize] [align=left] {\begin{minipage}[lt]{50.34835600000001pt}\setlength\topsep{0pt}
Probability
\begin{center}
space
\end{center}

\end{minipage}};
\draw (169.36,99.11) node [anchor=north west][inner sep=0.75pt]  [rotate=-269.26] [align=left] {Teacher};
\draw (169.79,193.93) node [anchor=north west][inner sep=0.75pt]  [rotate=-270.45] [align=left] {Student};
\draw (246,58.4) node [anchor=north west][inner sep=0.75pt]    {$\theta _{0}$};
\draw (384,66.57) node [anchor=north west][inner sep=0.75pt]    {$m_{0}$};
\draw (493,110.4) node [anchor=north west][inner sep=0.75pt]    {$p_{0}$};
\draw (228,156.4) node [anchor=north west][inner sep=0.75pt]    {$\theta '$};
\draw (373,156.4) node [anchor=north west][inner sep=0.75pt]    {$m'$};
\draw (247,9) node [anchor=north west][inner sep=0.75pt]   [align=left] {\textit{No communication}};
\draw (447,8) node [anchor=north west][inner sep=0.75pt]   [align=left] {\textit{Communication }};
\draw (320,60.4) node [anchor=north west][inner sep=0.75pt]    {$\mathcal{M}$};
\draw (314,158.4) node [anchor=north west][inner sep=0.75pt]    {$\mathcal{M} '$};
\draw (200,59.4) node [anchor=north west][inner sep=0.75pt]    {$\Theta $};
\draw (201,161.4) node [anchor=north west][inner sep=0.75pt]    {$\Theta '$};

\end{tikzpicture}

    \caption{The parameter space $\Theta'$ of the student can be different from the one  of the teacher, $\Theta$, and so can the model classes $\mathcal{M}$ and $\mathcal{M}'$. This prevents the teacher from communicating relevant information to the student. The space of induced probability distributions on the observable random variable ($y_k$) is, however, joint -- in this space, comprehensible communication can take place. The dashed arrow denotes a joint model class $\mathcal{M} = \mathcal{M}'$, but different parametrizations.} 
    \label{fig:spaces}
\end{figure}

\subsection{Means of communication}

Suppose that the system $m_0$ (and all models in $\mathcal{M}$) produces independent and identically distributed (i.i.d.) observations:\\

\begin{assumption}
    The observations $y_k$ produced by the true system $m_0$ (and the all other models in $\mathcal{M}$) are i.i.d.
    \label{ass:iid}
\end{assumption}

Then it is well known that, under suitable assumptions, the notion of identifiability of a model $m$ says that there are no two models that imply the same probability distribution on the observable random variable $y_k$ -- see, e.g., \cite{rothenberg1971identification}.

Hence, if the teacher and student are prohibited from directly transfering information in terms of model parameters (due to, e.g., privacy, bandwidth, parametrization), they can \emph{equivalently} operate in the space of induced probability distributions. We provide a visualization of this in Fig. \ref{fig:spaces}, where the student and teacher are unable to communicate with each other on the left side of the dashed line; because of, e.g., mismatched model classes ($\mathcal{M} \neq \mathcal{M}'$), mismatched parametrization ($\theta_0$ vs $\theta'$) or communication constraints. On the right side of the dashed line is the space of induced probability distributions over the observable random variable (i.e., the $y_k$'s) which is common to both the teacher and the student. 

Therefore, under Assumptions \ref{ass:model_class} and \ref{ass:iid}, and alluding to the identifiability of the model, the learning problem \eqref{eq:theta} can be reformulated in terms of the induced probability distribution of $y_k$. The student thus computes an empirical estimate\footnote{If the observation space $\mathcal{Y}$ is finite, then $\hat{p}$ corresponds to an empirical \emph{probability mass function} (pmf): $\hat{p}(y) = \frac{1}{N}\sum_{k=1}^N I(y_k = y)$ for $y \in \mathcal{Y}$, where $I$ denotes the indicator function; and if $\mathcal{Y}$ is a continuum, then a kernel density estimator (see \citet{parzen1962estimation}) can be used to compute $\hat{p}$.} $\hat{p}(y)$ and then tunes the parameters $\theta$ so as to minimize the discrepancy between the empirical distribution $\hat{p}$ and that induced by the model $p_\theta$:

\begin{equation}
\hat{\theta} \in \arg \min_{\theta \in \Theta} G(\hat{p}, p_\theta),
\label{eq:student_p}
\end{equation}

where $G$ is a distance measure between two probability density functions.

To improve the student's estimate, the teacher, who knows the true distribution $p_{\theta_0}(y)$ of the data, could correct some observations $y_k$ to $\tilde{y}_k$, with the aim of improving the student's empirical estimate $\hat{p}$. This serves as a way of side-stepping the requirement of a joint model class and/or parametrization -- if both parametrizations are identifiable, then having $\hat{p} = p_{\theta_0}$ will allow the student to reconstruct the true parameters even in its own parametrization.

\subsection{General correctional learning}
\label{sec:general_cl}

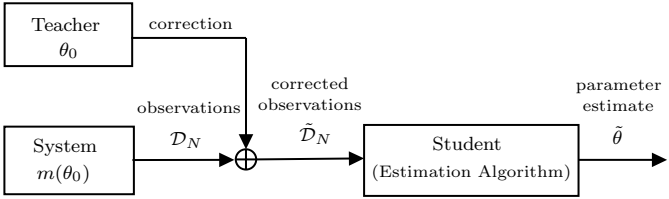
\begin{figure}[t]
\centering

\tikzset{every picture/.style={line width=0.75pt}} 

\begin{tikzpicture}[x=0.75pt,y=0.75pt,yscale=-0.85,xscale=0.85, every node/.style={scale=0.9}]

\draw    (251.33,101.77) -- (308.17,101.67) ;
\draw [shift={(311.17,101.67)}, rotate = 539.9] [fill={rgb, 255:red, 0; green, 0; blue, 0 }  ][line width=0.08]  [draw opacity=0] (8.93,-4.29) -- (0,0) -- (8.93,4.29) -- cycle    ;
\draw    (513.17,101.67) -- (562.17,101.67) ;
\draw [shift={(565.17,101.67)}, rotate = 180] [fill={rgb, 255:red, 0; green, 0; blue, 0 }  ][line width=0.08]  [draw opacity=0] (8.93,-4.29) -- (0,0) -- (8.93,4.29) -- cycle    ;
\draw    (318,29.5) -- (317.98,92.4) ;
\draw [shift={(317.97,95.4)}, rotate = 270.02] [fill={rgb, 255:red, 0; green, 0; blue, 0 }  ][line width=0.08]  [draw opacity=0] (8.93,-4.29) -- (0,0) -- (8.93,4.29) -- cycle    ;
\draw    (318,29.5) -- (250.67,29.5) ;
\draw   (311.97,101.4) .. controls (311.97,98.08) and (314.66,95.4) .. (317.97,95.4) .. controls (321.29,95.4) and (323.97,98.08) .. (323.97,101.4) .. controls (323.97,104.71) and (321.29,107.4) .. (317.97,107.4) .. controls (314.66,107.4) and (311.97,104.71) .. (311.97,101.4) -- cycle ; \draw   (311.97,101.4) -- (323.97,101.4) ; \draw   (317.97,95.4) -- (317.97,107.4) ;
\draw    (323.97,101.4) -- (384.3,100.83) ;
\draw [shift={(387.3,100.8)}, rotate = 539.46] [fill={rgb, 255:red, 0; green, 0; blue, 0 }  ][line width=0.08]  [draw opacity=0] (8.93,-4.29) -- (0,0) -- (8.93,4.29) -- cycle    ;
\draw   (176.17,82) -- (252.17,82) -- (252.17,122) -- (176.17,122) -- cycle ;
\draw   (176.3,9) -- (251.17,9) -- (251.17,46.17) -- (176.3,46.17) -- cycle ;
\draw   (387.3,81) -- (513.17,81) -- (513.17,123) -- (387.3,123) -- cycle ;

\draw (212,94) node  [font=\small,color={rgb, 255:red, 0; green, 0; blue, 0 }  ,opacity=1 ] [align=left] {System};
\draw (212.3,21.47) node  [font=\small,color={rgb, 255:red, 0; green, 0; blue, 0 }  ,opacity=1 ] [align=left] {Teacher};
\draw (211,110) node  [font=\small]  {$m( \theta _{0})$};
\draw (359,86) node  [font=\small]  {$\tilde{\mathcal{D}}_{N}$};
\draw (283,90) node  [font=\small]  {$\mathcal{D}_{N}$};
\draw (284.46,70.67) node  [font=\scriptsize,color={rgb, 255:red, 0; green, 0; blue, 0 }  ,opacity=1 ] [align=left] {observations};
\draw (214,37) node  [font=\small]  {$\theta _{0}$};
\draw (285.46,20.67) node  [font=\scriptsize,color={rgb, 255:red, 0; green, 0; blue, 0 }  ,opacity=1 ] [align=left] {correction};
\draw (355.46,62.67) node  [font=\scriptsize,color={rgb, 255:red, 0; green, 0; blue, 0 }  ,opacity=1 ] [align=left] {\begin{minipage}[lt]{43.208356pt}\setlength\topsep{0pt}
\begin{center}
corrected\\observations
\end{center}

\end{minipage}};
\draw (448,92) node  [font=\small,color={rgb, 255:red, 0; green, 0; blue, 0 }  ,opacity=1 ] [align=left] {Student};
\draw (450.23,109.15) node  [font=\scriptsize,color={rgb, 255:red, 0; green, 0; blue, 0 }  ,opacity=1 ] [align=left] {(Estimation Algorithm)};
\draw (536,87) node  [font=\small]  {$\tilde{\theta }$};
\draw (536.46,60.67) node  [font=\scriptsize,color={rgb, 255:red, 0; green, 0; blue, 0 }  ,opacity=1 ] [align=left] {\begin{minipage}[lt]{35.267928000000005pt}\setlength\topsep{0pt}
\begin{center}
parameter\\estimate
\end{center}

\end{minipage}};

\end{tikzpicture}

\caption{Overview of the setup in cooperative system identification using correctional learning. A student aims to estimate a model of a system from observed data. An expert (teacher) is able to intercept and modify the observations to accelerate the learning process \eqref{eq:min_goodness_of_fit_model}.}
\label{fig:generalSysId}
\end{figure}

A crucial question is: \emph{Which observations should the teacher modify, and to what?} 
We provide a schematic illustration of the correctional learning problem in Fig.  \ref{fig:generalSysId}, and give a formal statement below.\\

\begin{problem}(Correctional Learning for Cooperative System Identification).
Under Assumptions \ref{ass:model_class} and \ref{ass:iid}, a set of $N$ observations $\mathcal{D} = \{y_k\}_{k=1}^N$ has been sampled from a system, as in \eqref{eq:observations}. 
A teacher, who knows the true system $m_0$, and hence also the true probability distribution $p_{\theta_0}$ of the observations, is able to modify the data $\mathcal{D}$ before the student receives it. Provide a method for the teacher to improve the learning process of the student, who is acting according to problem \eqref{eq:min_goodness_of_fit_model}, by modifying $\mathcal{D}$. 
\label{pr:cl4sysid}
\end{problem}

Denote the modified set of observations as $\tilde{\mathcal{D}} = \{ \tilde{y}_k \}_{k=1}^N$. Realistically, it is reasonable to assume that the teacher has a limited budget $b \in \mathbb{R}_0^+$ (e.g., if modifying radar measurements, there would be a budget on power that would reveal its inference). Let $B(\mathcal{D}, \tilde{\mathcal{D}})$ be a function that measures the distance between the two sets $\mathcal{D}$ and $\tilde{\mathcal{D}}$. This could, for example, be the $\ell_0$-norm that computes the number of changed observations, or a more sophisticated measure that weighs larger modifications of observations heavier than smaller ones. 

We also introduce a function $V(p, \tilde{p})$ that measures the discrepancy between two probability distributions. In terms of Problem \ref{pr:cl4sysid}, $\tilde{p}$ would be the empirical distribution of the observations $y_k$ computed using the modified dataset $\tilde{\mathcal{D}}$ -- i.e., the empirical estimate computed by the student. The function $V$ would generally take into account the measure of fit that the student uses to compute a parameter estimate, see equation \eqref{eq:student_p}.  However, if the teacher lacks this knowledge, then a general measure (such as the Kullback-Leibler divergence) can be employed. 

Since the goal in Problem \ref{pr:cl4sysid} is to minimize the distance between the empirical distribution computed by the student and the true distribution, a solution to this problem is formulated as an optimization problem as:\\
\begin{equation}
\begin{aligned}
\min_{\tilde{\mathcal{D}}  }  \quad & V(p_0, \tilde{p}) \\
\text{s.t.} \quad & \tilde{y}_k \in \mathcal{Y}, \text{ for all } \tilde{y}_k \in \tilde{\mathcal{D}},\\
 &  B( \mathcal{D} , \tilde{\mathcal{D}} ) \leq b.\\
\end{aligned}
\label{eq:batchopt}
\end{equation}



\begin{remark}
Note that problem \eqref{eq:batchopt} has strong connections to the optimal mass transport problem -- e.g., \cite{kolouri_optimal_2017}. Below, we provide a direct solution, but in future work it would be of interest to explore how more specialized algorithms can be used.
\end{remark}


\section{Cooperative System Identification for Bi- and Multinomial Systems}
\label{sec:examples}

In the previous section, we gave a general solution \eqref{eq:batchopt} to the cooperative system identification problem (Problem \ref{pr:cl}) using correctional learning (Problem \ref{pr:cl4sysid}). In this section, we make the general optimization problem \eqref{eq:batchopt} concrete by considering two specific systems: multinomial and binomial. Albeit structurally simple, they help make the terms used in problem \eqref{eq:batchopt} tangible, and allow us to derive insightful theoretical finite-sample bounds (that could later be used as a basis for generalizations).

\subsection{Solution for multinomial systems}
\label{ssec:multinomial}

Suppose that each observation $y_k \in \mathcal{Y} = \{1, \dots, Y\}$ has been sampled i.i.d. according to probabilities $[\theta_0]_i = p_0(y_k = i).$\footnote{Note that we adopt a trivial parametrization: $\theta_0 = p_0$.} Note that this means that $p_0$ and $\tilde{p}$ (the empirical distribution computed by the student) are $Y$-dimensional probability mass functions. We identify these with vectors in $\mathbb{R}^Y$. 

For simplicity, we adopt the $\ell_2$-norm 
as distance measure $V$ and the $\ell_1$-norm as cost measure $B$. Moreover, assume that the student employs a standard maximum-likelihood criterion for its estimation \eqref{eq:min_goodness_of_fit_model}. Then, introducing this in \eqref{eq:batchopt} yields the problem
\begin{equation}
\begin{aligned}
\min_{\tilde{\mathcal{D}}  }  \quad & \| p_0 \; - \; \tilde{p} \|_2 \\   
\text{s.t.} \quad & \tilde{y}_k \in \mathcal{Y}, \text{ for all } \tilde{y}_k \in \tilde{\mathcal{D}},\\
& [\tilde{p}]_i = \frac{1}{N} \sum_{k=1}^N I( \tilde{y}_k = i ), \\ 
& \hspace{1cm} \text{ for all } i \in \{1, \dots, Y\},\\
& \sum_{k=1}^N | y_k - \tilde{y}_k | < b,
\end{aligned}
\label{eq:batchopt_specific}
\end{equation}
where $y_k$ and $\tilde{y}_k$ are elements of $\mathcal{D}$ and $\tilde{\mathcal{D}}$, respectively, and $b \in \mathbb{R}^+$.
To translate \eqref{eq:batchopt_specific} into a standard integer program, we define the $Y \times N$ matrix
\begin{equation}
    [D]_{ij} = I(y_j = i),
\end{equation}
where $I$ denotes the indicator function. Matrix $D$ has a one on row $i$ and column $j$ if observation $i$ was sampled at time $j$. With this reparametrization, solving Problem \ref{pr:cl4sysid} for a multinomial distribution thus corresponds to finding the matrix $\tilde{D}$ with the same shape as $D$ but for the altered observations.

\begin{equation}
\begin{aligned}
    \min_{\tilde{D} \in \mathbb{R}^{Y\times N}} & \quad \| p_0 \; - \; \tilde{p} \|_2   \\  
    \text{s.t.} & \quad [\tilde{D}]_{ij} \in \{0,1\} \text{ for all } i, j, \\ 
    & \quad  \tilde{{p}} = \frac{1}{N} \tilde{D} \mathds{1}    , \\
    & \quad \mathds{1}^T \tilde{D} = \mathds{1}^T, \\
    & \quad  \frac{1}{2} \| \textnormal{vec}(D - \tilde{D})  \|_1 \leq b.
\end{aligned}
\label{eq:multinomial}
\end{equation}
In the optimization problem \eqref{eq:multinomial}, the term $\frac{1}{2}$ derives from each column of the observations matrices $D$ and $\tilde{D}$ having two different values when the corresponding observation is changed. 
This problem can be solved using standard off-the-shelf solvers such as Mosek, Gurobi and IBM CPLEX. 
To recover the sequence of altered observations, one can use the relation
\begin{equation}
    \tilde{y}_k = \left[ \begin{bmatrix} 1 & \dots & Y \end{bmatrix} \tilde{D} \right]_{k}, \quad k=1,\dots,N
\end{equation}



\subsection{Finite-sample results for binomial systems}
\label{ssec:binomial}

In order to derive finite-sample results, we will now consider a special case of the multinomial distribution with only two possible observations: the binomial.
Each observation $y_k \in \mathcal{Y} = \{0, 1\}$ is now sampled i.i.d. according to a Bernoulli distribution:
\begin{equation}
    \theta_0 = p_0(y_k = 1) = 1 - p_0(y_k = 0).
\end{equation} 

Let $X$ be the number of positive outcomes (``successes'') that have been observed in $N$ trials in the dataset $\mathcal{D}$:
\begin{equation}
    X = \sum_{k=1}^N y_k.
\end{equation}
Its probability mass function is
\begin{equation}
p_0(X=x) = \left(\begin{array}{l}
N \\ x \end{array}\right) (\theta_0)^{x}(1-\theta_0)^{N-x},
\label{eq:binomial}
\end{equation}
and expected value $\mathbb{E}[X] = N\theta_0$ -- see, e.g., \citep{papoulis2002probability}.

Our main theoretical results are the following: \\



\begin{theorem}   
    Consider a binomial process with success probability $\theta$. There are $N$ realized outcomes $\mathcal{D} = \{y_1, \dots, y_N\}$. For simplicity, assume that $\mathbb{E}[X] =N \theta_0$ is an integer. Let $\tilde{\mathcal{D}} = \{\tilde{y}_1, \dots, \tilde{y}_N\}$ be the dataset corrected by the teacher with budget $b$. The number of successes in the altered observations is denoted $\tilde{X}$ and its probability mass function is:
   \begin{align}
       p(\tilde{X}=\tilde{x}) = \begin{cases}
       p(X=\tilde{x}-b), \quad & \textnormal{if } \tilde{x}<\mathbb{E}[X],\\
       p(X=\tilde{x}+b), \quad & \textnormal{if } \tilde{x}>\mathbb{E}[X], \\
       \sum_{b'=-b}^{b} p(X=\mathbb{E}[X]+b'), \quad & \textnormal{if } \tilde{x}=\mathbb{E}[X],\\
       \end{cases}
       \label{eq:altered_probabilities}
   \end{align}
   considering that $p(X<0)=p(X>N)=0$.\footnote{Otherwise, one can add the condition $\tilde{p}=0 \textnormal{ if } (\tilde{x} \leq b \textnormal{ or } \tilde{x}> N-b$).}
   \label{thm:probabilities}
\end{theorem}
\begin{pf}(outline)
For the binomial case, it is easy to heuristically solve problem \eqref{eq:multinomial}: one simply ``changes'' as many observations as the budget permits (the order is unimportant). Note, however, that this is not the case for more general systems (e.g., the multinomial). 
\end{pf}

With the probability mass function characterized, we can now derive and analyze the properties of the new estimator, such as its expected value and variance:\\
\begin{theorem} 
     Let $\hat{\theta}$ be the estimate computed using the uncorrected dataset $\mathcal{D}$, and $\tilde{\theta}$ the estimate computed using the corrected one, $\tilde{\mathcal{D}}$:
    \begin{equation}
        \tilde{\theta} = \frac{1}{N} \sum_{k=1}^N \tilde{y}_k.
    \end{equation}
    Then, the variance of $\tilde{\theta}$ is given by
    \begin{align}
        \textnormal{var}\{\tilde{\theta}\} = \textnormal{var}\{\hat{\theta}\} - \delta(N,b,p_0),
        \label{eq:variance_altered}
    \end{align}
    where $\delta(N,b,p_0) \geq 0$ for all $b$ (with equality when $b = 0$) -- that is, $\textnormal{var}\{\tilde{\theta}\} \leq \textnormal{var}\{\hat{\theta}\}$.
    \label{thrm:variance_corrected_binomial}
\end{theorem}
\begin{pf}(outline)
The expression for the variance is obtained by replacing the probability mass function \eqref{eq:altered_probabilities} in the definition: $\textnormal{var}\{\tilde{\theta}\} = \frac{1}{N^2}(\mathbb{E}[\tilde{X}^2]- \mathbb{E}[\tilde{X}]^2)$, where $\mathbb{E}[\tilde{X}]=\sum_{\tilde{x}} \tilde{x} \; p(\tilde{X}=\tilde{x})$ and $\mathbb{E}[\tilde{X}^2] = \sum_{\tilde{x}} \tilde{x}^2 \; p(\tilde{X}=\tilde{x})$.
Algebraic manipulations are available in Appendix \ref{app:binomial}, together with the expression for $\delta(N,b,p_0)$.
\end{pf}
Theorem \ref{thrm:variance_corrected_binomial} guarantees that the teacher is, in fact, helping the student (by reducing the variance of its estimator).
With the variance of the modified estimator computed, we can now compare it to the ``uncorrected'' estimator:\\

\begin{corollary}
    As the budget of the teacher increases, $b \to N$, we have that 
    \begin{equation}
        \frac{\textnormal{var}\{\tilde{\theta}\}}{\textnormal{var}\{\hat{\theta}\}} \to 0,
        \label{eq:variance_comparison}
    \end{equation}
    \label{cor:ratio_of_variances}
    for $0<\theta_0<1$.\footnote{If $\theta_0$ is 0 or 1, then all observations are equal and $\text{var}(\hat{\theta}) = 0$.}
\end{corollary}
\vspace{-0.2cm}
\begin{pf}(outline)
First, the variance of $\hat{\theta}$ is given by
\begin{equation}
\textnormal{var}\{\hat{\theta}\}= \theta_0(1-\theta_0)/N,
\label{eq:variance_original}
\end{equation} 
(see, e.g., \cite{papoulis2002probability}), which is constant in $b$, and the variance of the modified estimator $\tilde{\theta}$ is given in \eqref{eq:variance_altered}. 
For $b=N$, the probability $\tilde{p}$ from \eqref{eq:altered_probabilities} is one if $\tilde{x}=\mathbb{E}[X]$, and zero otherwise. Therefore, for $b=N$, the variance of $\tilde{\theta}$ is zero. Since the denominator of $\eqref{eq:variance_comparison}$ is constant in $b$, the ratio is zero. More details are given in Appendix \ref{app:binomial}.
\end{pf}

The importance of Corollary~\ref{cor:ratio_of_variances} is that as the budget increases, the teacher is able to increasingly accurately transfer its knowledge to the student -- that is, to successfully perform cooperative system identification. The quantity in equation \eqref{eq:variance_comparison} is a measure of the improvement (in terms of variance reduction compared to the original estimator) that can be obtained using correctional learning. 




\section{Numerical results}
\label{sec:results}

In this section we validate the theoretical results presented in the previous sections in numerical experiments. 
All simulations were implemented in Python 3.7 and run on a 1.90 GHz CPU.

\subsection{Reduction in variance using correctional learning}
\label{ssec:coin}


Recall that in Problem \ref{pr:cl4sysid} (Section~\ref{sec:general_cl}) the teacher aims to transmit its knowledge of $p_0$ to the student by modifying the observations that the student sees. For a binomial distribution (Section~\ref{ssec:binomial}) with parameter $\theta_0 = 0.4$, we solve optimization problem \eqref{eq:multinomial}. Since the maximum-likelihood estimator is consistent for this setup, when $N~\rightarrow~\infty$, the student's estimator will converge to the true parameters (with probability one) even for an empty budget ($b=0$). 
However, the teacher aims to improve the student's estimate for finite data by reducing its variance.

\begin{figure}[t!]
    \centering
\begin{tikzpicture}
\begin{axis}[%
width=2.5in,
height=0.8in,
at={(0.758in,0.481in)},
scale only axis,
xmin=5,
xmax=30,
xlabel style={font=\color{white!15!black}},
xlabel={Number of observations, $N$},
ymin=0,
ymax=0.05,
ylabel style={font=\color{white!15!black}},
ylabel={var\{$\tilde{\theta}$\}},
axis background/.style={fill=white},
title style={font=\bfseries},
align =center, 
title={Reduction in Variance \\ Using Correctional Learning \\[-0.1cm]},
axis x line*=bottom,
axis y line*=left,
legend style={at={(0.7,1.2)}, anchor=north west, legend cell align=left, align=left, legend plot pos=left, draw=black}
]
\addplot[color=mycolor1, line width=1.0pt] table[col sep=comma, x=N,y=b0] {variance_binomial.txt};
\addplot[color=mycolor3, line width=1.0pt] table[col sep=comma, x=N,y=b1] {variance_binomial.txt};
\addplot[color=mycolor2, line width=1.0pt] table[col sep=comma, x=N,y=b2] {variance_binomial.txt};

\addplot[ color=mycolor1, dashed, line width=0.5pt] table[col sep=comma, x=even_n,y=b0t] {variance_binomial_theor.txt};
\addplot[ color=mycolor3, dashed, line width=0.5pt] table[col sep=comma, x=even_n,y=b1t] {variance_binomial_theor.txt};
\addplot[ color=mycolor2, dashed, line width=0.5pt] table[col sep=comma, x=even_n,y=b2t] {variance_binomial_theor.txt};
\legend{$b=0$, $b=1$, $b=2$}
\end{axis}
\end{tikzpicture}%
    \caption{Variance of the estimator for different budgets $b$ as $N$ increases. Each color represents a different budget, with the dashed being the theoretical (in \eqref{eq:variance_altered}) and the solid the experimental curves. The case $b = 0$ corresponds to an unmodified estimator (in \eqref{eq:variance_original}).}
    \label{fig:binomial_variance}
\end{figure}
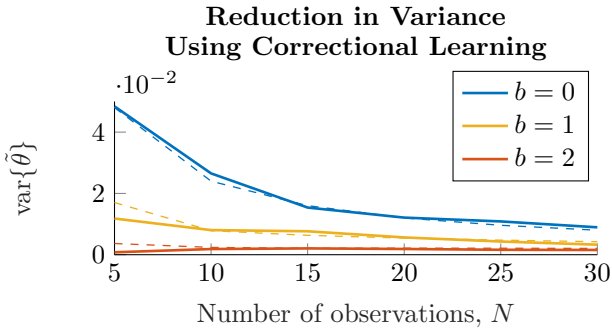

Using 200 Monte-Carlo simulations, we estimated the variance of the original estimator (blue curve) and that computed using correctional learning  (yellow and red curves).
Fig. \ref{fig:binomial_variance} demonstrates how the variance decreases as the budget increases. In dotted lines are plotted the corresponding theoretical curves from Theorem \ref{thrm:variance_corrected_binomial}. This figure demonstrates a good correspondence between the theoretical results and the numerical experiments. Note that with $b = 2$, i.e., when the teacher can modify two observations, the variance is reduced by up to an order of magnitude.


\section{Conclusions}
\label{sec:conclusions}



In this paper, we have introduced a framework for cooperative system identification. 
We proposed \emph{correctional learning} as novel method to solve this problem. In it, the teacher modifies the dataset of the student in order to make the empirical data distribution more close to the true system. This allows communication to take place between the teacher and student, even if they employ different model classes and/or parametrizations (which would otherwise hinder communication).
We derived variance bounds for the binomial distribution, which quantify how much the teacher can help the student. These bounds were implemented and compared against estimates in numerical experiments, showing a good correspondence.

\subsection{Future work}
\label{sec:others}

In future work, it would be of interest to investigate how correctional learning can be performed in an online setting where observations are received sequentially and the teacher has to decide at each time step whether or not to change it (and to what).
Moreover, we would like to study generalizations of the cooperative system identification problem (which aims to estimate static parameters) to, e.g., filtering (dynamic state estimation), the reinforcement learning (policy learning) and social learning.

\bibliography{correctional}             












\appendix
\addcontentsline{toc}{section}{Appendices}
\section*{Appendices}

\section{Influence of the budget}
\label{app:appendixA}

Appendix \ref{app:appendixA} presents results concerning the influence of the budget of the teacher in the estimation of the student, once the original sequence of observations, $D$, is received.

\subsection{Finite-sample results }

For the case of finite distributions such as the multinomial and binomial from Section \ref{sec:examples}, we can derive the following results when the distance measure $V$ from \eqref{eq:batchopt} is the $\ell_1$-norm:\\

\begin{lemma}[Minimum error]
The smallest error possible to attain, as a function of the number of observations $N$ and the true distribution, $p_0$, is:
\begin{equation}
e_{min}(N,p_0) = \|p_0- \frac{[p_0 N]}{N}\|_1,
\label{eq:error_min}
\end{equation}
where $[\cdot]$ means rounding to the closest integer value, and the rounding has to take into account the constraint $\mathds{1}^T \tilde{p} =  \mathds{1}^T$, where $\tilde{p} = \frac{[p_0 N]}{N}$. \\
\label{lem:emin}
\end{lemma}

\begin{lemma}[Minimum budget]
The minimum budget $b\in \mathbb{N}_0^+$ needed to attain the smallest error $e_{min}$, computed in \eqref{eq:error_min}, for a certain set of observations, is given by
\begin{equation}
b_{min}(N,p_0,\hat{p})=  \lceil \frac{N}{2} ( \| p_0 - \hat{p} \|_1 - e_{min}) \rceil, 
\end{equation}
where $\hat{p}$ is the distribution of the original data, $\tilde{p}$ is the distribution of the altered data, and $\lceil \cdot \rceil$ is the ceiling function. \\
\label{lem:budgetmin}
\end{lemma}

\begin{theorem}[Estimation error]
The estimation error is the difference between the corrected parameter, $\tilde{\theta}$, and the true one, $\theta_0$, and is given by a function of the number of observations $N$, the true distribution $p_0$, the budget $b$ and the original sequence $\hat{p}$, as:
\begin{equation}
e(N,p_0,b,\hat{p}) = \max\{ \| p_0 - \hat{p} \|_1 - \frac{2 b}{N} , e_{min} \}.\\
\end{equation}
\label{thm:error}
\end{theorem}

\subsection{Numerical experiment}

For the scenario of the multinomial distribution presented in Section \ref{ssec:multinomial}, we solve the optimization problem \eqref{eq:multinomial} in the case where $Y=3$ and the distribution has true parameter $p_0=[0.12,0.63,0.25]^T$. The empirical estimate computed using the unmodified dataset with $N=17$ observations was $\hat{p}=[0.2,0.5,0.3]^T$. 

\begin{figure}[ht!]
    \centering
\begin{tikzpicture}

\begin{axis}[%
width=2.5in,
height=1.5in,
at={(0.758in,0.481in)},
scale only axis,
xmin=0,
xmax=6,
xlabel style={font=\color{white!15!black}},
xlabel={Budget, $b$},
ymin=0,
ymax=0.3,
ylabel style={font=\color{white!15!black}},
ylabel={Error},
ytick={0.26,0.16,0.04},
axis background/.style={fill=white},
title style={font=\bfseries},
title={Estimation Error},
axis x line*=bottom,
axis y line*=left,
legend style={at={(0.7,0.77)}, anchor=north west, legend cell align=left, align=left, legend plot pos=left, draw=black}
]
\addplot[const plot, color=mycolor2, line width=1.0pt] table[col sep=comma, x=cb,y=e_hat] {error_multinomial.txt};
\addplot[const plot, color=mycolor3, line width=1.0pt] table[col sep=comma, x=cb,y=e_bar] {error_multinomial.txt};
\addplot[const plot, black, dashed, line width=0.5pt] table[col sep=comma, x=cb,y=e_bar] {error_multinomial.txt};
\addplot[only marks, mark=x, mark options={}, mark size=2.000pt, draw=mycolor4, forget plot] table[col sep=comma, x=cb,y=e_hat] {error_multinomial.txt};
\addplot[only marks, mark=x, mark options={}, mark size=2.000pt, draw=mycolor4, forget plot] table[col sep=comma, x=cb,y=e_bar] {error_multinomial.txt};
\addplot [domain = 0:6, dashed, ]{0.04};
\addplot +[mark=none, dashed, black] coordinates {(3, 0) (3, 0.6)};
\draw (340,290) node  [font=\small,color={rgb, 255:red, 0; green, 0; blue, 0 }  ,opacity=1 ] [align=left] {$b_{min}$};
\draw (50,50) node  [font=\small,color={rgb, 255:red, 0; green, 0; blue, 0 }  ,opacity=1 ] [align=left] {$e_{min}$};
\legend{$\|p_0-\hat{p}\|_1$, $\|p_0-\tilde{p}\|_1$}
\end{axis}
\end{tikzpicture}%
    \caption{Evolution of the estimation errors as the budget increases. Without the teacher, the student has an estimation error of $\|p_0-\hat{p}\|_1=0.26$, and with the help of the teacher this error decreases to $\|p_0-\tilde{p}\|_1 = 0.04$ for any budget bigger than $b=3$. The theoretical results are marked in dashed lines and coincide with the experimental ones.}
    \label{fig:multinomial_budget}
\end{figure}
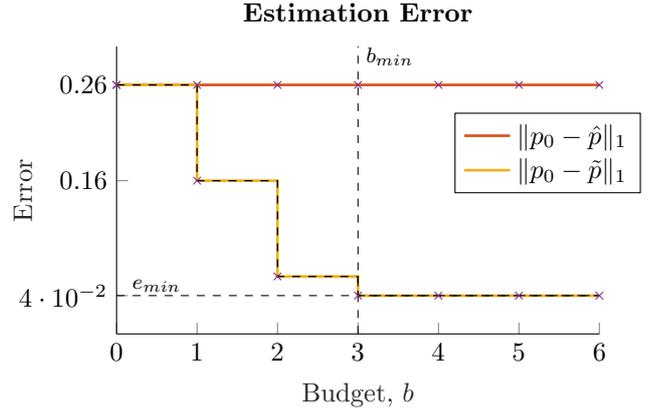

Fig. \ref{fig:multinomial_budget} shows the estimation error of the student without the teacher, in red ($\|p_0-\hat{p}\|_1=0.26$), and how it decreases with the help of the teacher, in yellow, as its budget increases. From the figure we see that the minimum error is ($\|p_0-\tilde{p}\|_1 \approx 0.04$), which is obtained for a minimum budget of $b=3$. The corresponding theoretical values computed from Lemmas \ref{lem:emin} and \ref{lem:budgetmin} are plotted in dashed lines, and coincide with the experimental ones. The same is true for the error expression computed in Theorem \ref{thm:error}.

\section{Binomial finite-sample results}
\label{app:binomial}

Appendix \ref{app:binomial} presents the full expression of the variance presented in Theorem \ref{thrm:variance_corrected_binomial}, as well as more detailed proofs of this theorem and Corollary \ref{cor:ratio_of_variances}.

\subsection{Expression for the variance of $\tilde{\theta}$}
The variance is defined as $\textnormal{var}\{\tilde{\theta}\} = \frac{1}{N^2}(\mathbb{E}[\tilde{X}^2]- \mathbb{E}[\tilde{X}]^2)$, where by replacing the probability mass function \eqref{eq:altered_probabilities} we obtain

\begin{equation}
\begin{aligned}
    \mathbb{E}[\tilde{X}] &= \sum_{\tilde{x}=0}^{\mathbb{E}[X]-1} \tilde{x} \; p(X=\tilde{x}-b) \\
    &+ \sum_{\tilde{x}=\mathbb{E}[X]+1}^{N} \tilde{x} \; p(X=\tilde{x}+b) \\
    &+ \sum_{b'=-b}^{b} \mathbb{E}[X] \; p(X=\mathbb{E}[X]+b'),
\end{aligned}
\end{equation}
and similarly for $\mathbb{E}[\tilde{X}^2]$. 
By manipulating the expression resulting from the subtraction of the expected values and diving by $\frac{1}{N^2}$ we obtain
 \begin{align}
    \textnormal{var}\{\tilde{\theta}\} = \textnormal{var}\{\hat{\theta}\} + \delta(N,b,p_0),
    \end{align}
    where 
    
\begin{equation}
    \begin{aligned}
    \delta(N,b&,p_0) =\\
     \frac{1}{N^2} \Bigg[ & -\sum_{k=-1:b}^{b-1} \big(k^2 + 2k\mathbb{E}[X]\big)  p(X = \mathbb{E}[X]+k)\\
     &+ b^2 \; \big(\textnormal{cdf}(\mathbb{E}[X]-b-1) - \textnormal{cdf}(\mathbb{E}[X]+b) +1\big) \\
    &+ 2b \; \bigg( \sum_{z=0}^{\mathbb{E}[X]-b-1}z \; p(X=z) - \sum_{z=\mathbb{E}[X]+b+1}^{N}z \; p(X=z) \bigg)\\
    &- \phi^2  - 2 \phi \mathbb{E}[X] \Bigg],
\end{aligned}
\end{equation}
and 

\begin{equation}
\begin{aligned}
    \phi =&  \sum_{z=-b}^{b} z \; p(X= \mathbb{E}[X] +z) \\
    +& \; b \; \big(\textnormal{cdf}(\mathbb{E}[X]-b-1) + \textnormal{cdf}(\mathbb{E}[X]+b) -1\big).
\end{aligned}
\end{equation}

\subsection{The term $\delta(N,b,p_0)$ is non-positive}

First, note that if $b=0$, $\delta(N,b,p_0) = 0$ and therefore $\textnormal{var}\{\tilde{\theta}\} = \textnormal{var}\{\hat{\theta}\}$.

We now show that $\delta(N,b,p_0) \leq 0$.
\begin{itemize}
    \item The first term is negative since the probabilities and their multiplicative factors are positive. 
    \item The second term can be rewritten as
    \begin{equation}
\begin{aligned}
    &b^2 \; \big(\textnormal{cdf}(\mathbb{E}[X]-b-1) - \textnormal{cdf}(\mathbb{E}[X]+b) +1\big) \\ 
    = \; &b^2 - b^2 \sum_{z=\mathbb{E}[X]-b}^{\mathbb{E}[X]+b} P(x=z),
\end{aligned}
    \end{equation}
where the only positive term is $b^2$. However, this term will be cancelled with the $-b^2$ term resulting from $-\phi^2$.
\item The third term is positive but is smaller than the second term of $\phi$, which is negative since it is obtained from $-2\mathbb{E}[X]\phi$:
\begin{equation}
\begin{aligned}
    \Bigg| 2b \; \bigg( \sum_{z=0}^{\mathbb{E}[X]-b-1}z \; p(X=z) - \sum_{z=\mathbb{E}[X]+b+1}^{N}z \; p(X=z) \bigg) \Bigg|\\
    < \big| 2b \mathbb{E}[X] \; \big(\textnormal{cdf}(\mathbb{E}[X]-b-1) + \textnormal{cdf}(\mathbb{E}[X]+b) -1\big)  \big|.
\end{aligned}
\end{equation}
Using the fact that cdf$(z)= \sum_{z=0}^{N} \; p(X=z)$, 
the sum of the two terms on the left is 
\begin{align}
    2b\sum_{z=0}^{\mathbb{E}[X]-b-1} (z-\mathbb{E}[X])  \; p(X=z)
\end{align}
which is $<0$ since $z<\mathbb{E}[X]$, and similarly for the ones on the right.
\end{itemize}
Since all positive terms in the expression of $\delta(N,b,p_0)$ are smaller than certain negative ones, we conclude that the sum of all terms is negative, therefore $\delta(N,b,p_0) \leq 0$.

\subsection{The variance of $\tilde{\theta}$ goes to zero faster than that of $\hat{\theta}$}

We finally show that the variance of $\tilde{\theta}\to 0$ when $b\to N$ (Corollary \ref{cor:ratio_of_variances}).
From \eqref{eq:altered_probabilities}, with $b=N$ we obtain
\begin{align}
   p(\tilde{X}=\tilde{x}) = \begin{cases}
   p(X=\tilde{x}-N), \quad & \textnormal{if } \tilde{x}<\mathbb{E}[X],\\
   p(X=\tilde{x}+N), \quad & \textnormal{if } \tilde{x}>\mathbb{E}[X], \\
   \sum_{b'=-N}^{N} p(X=\mathbb{E}[X]+b'), \quad & \textnormal{if } \tilde{x}=\mathbb{E}[X].
   \end{cases}
\end{align}
The first probability becomes zero since if $\tilde{x}<\mathbb{E}[X]$, $\tilde{x}-N<0$ and we considered that $p(X<0)=0$.
Similarly for the second term, since if $\tilde{x}>\mathbb{E}[X]$, $\tilde{x}+N>N$ and we considered that $p(X>N)=0$.
We are left with the last term, that can be re-written as 
\begin{align}
\sum_{z=\mathbb{E}[X]-N}^{\mathbb{E}[X]+N} p(X=z), \quad  \textnormal{if } \tilde{x}=\mathbb{E}[X].
\end{align}
The limits of the summation can be rewritten from 0 to $N$, since for all other values the probability would be zero as previously mentioned. Therefore, the final probability density function is given by
\begin{align}
p(\tilde{X}=\tilde{x}) = \begin{cases}
\sum_{z=0}^{N} p(X=z)=1, \quad &\text{if } \tilde{x}=\mathbb{E}[X],\\
0, \quad &\textnormal{otherwise}.
\end{cases}
\end{align}
The variance is zero since the probability is all concentrated in one outcome.

Since var$(\hat{\theta})\neq 0$ for $0<\hat{\theta}<1$, the ration of both variances in \eqref{eq:variance_comparison} is zero.
\end{document}